# Formation of fluorescent H – aggregates of a cyanine dye in ultrathin film and its effect on energy transfer.

Santanu Chakraborty, Pintu Debnath, Dibyendu Dey, D. Bhattacharjee, Syed Arshad Hussain*

Thin Film and Nanoscience Laboratory, Department of Physics, Tripura University (A Central University), Suryamaninagar – 799022, Tripura, India

*Corresponding author
Email: sa_h153@hotmail.com, sahussain@tripurauniv.in
Phone: +919862804849 (M); +913812375317 (O)
Fax: +913812374802 (O)

**Abstract**

In this communication we report the formation of fluorescent H – aggregates of a cyanine dye 3,3 – Dioctadecyloxacarbocyanine perchlorate (Oxa18) in ultrathin film and its effect on energy transfer between Oxa18 and Sulphorhodamine B monosodium salt (sRhb). Surface pressure – area per molecule isotherm revealed that Oxa18 forms stable Langmuir monolayer at air-water interface. Spectroscopic investigation revealed that Oxa18 forms H – aggregate in Langmuir – Blodgett (LB) film which is enhanced in presence of nano clay laponite. Ideally H – aggregate do not fluoresce. However, due to imperfect stacking of Oxa18 molecules in the aggregates, fluorescence occurred from Oxa18 H – aggregates. This H – aggregated band has substantial effect on the enhancement of energy transfer from Oxa18 to sRhb both in solution and in ultrathin film.

# 1. Introduction

The dependence of optical and physicochemical behaviors of molecular assemblies upon the molecular arrangement is the current topic of research [1 - 3]. Recently a wide range of nanostructures and ultrathin films having various properties relating to the application in different fields have been developed [2 - 3]. It is well known that molecular structure not only determines its properties but also its patterns when assembled onto ultrathin films. Langmuir–Blodgett (LB) technique is one of the most promising candidates for fabricating nano-scale systems by arranging various kinds of molecules onto uniform monolayer assemblies, which may be suitable for development of different sized molecular aggregates with variety of functionality [1 – 4].

Depending on the molecular orientation in the aggregate, J – aggregate and H – aggregate are formed. In J – aggregate the molecules are aligned in a head to tail arrangement. J – aggregates are characterized by a sharp band, red shifted with respect to the monomer and by a strong photoluminescence with almost zero stokes shift. On the other hand in H – aggregate molecular alignment is side – by – side. This type of aggregate is ideally non fluorescent in nature [5]. Normally such aggregation behavior of fluorescent dye molecules strongly quenches their fluorescence, which has adverse effect on designing display devices and sensors as this applications are mainly based on their fluorescence properties. So it is very much important for the applicability of these fluorescent dyes either to control aggregation or to get fluorescence from these aggregates. Aggregation of dye molecules can be controlled by mixing these dyes with different fatty acid matrices. But in fatty acid matrix the presence of pure dye component in the film decreases which may effects the overall performance of the particular device. In this respect fluorescent behavior of H – aggregate is very much interesting. It is relevant to mention in this context that fluorescence behavior of H – aggregates of some cyanine derivatives have already been reported [6-9]. However, this rare H – aggregate fluorescence is also observed in other dye types [10-12]. The most important and unlikely behavior is seen in case of the dye coumarin 481, where the fluorescence intensity of H – aggregates unexpectedly exceeds the monomer fluorescence intensity [10].

H-aggregate can effectively contribute to the photocurrent in Schottky-type photovoltaic cells where thin films are sandwiched between two different types of metal electrodes [13]. Also H – aggregates can be used to study non linear optical phenomena as strong coherent excitation phenomena of H – aggregates show higher optical coefficients for third order nonlinear susceptibility [14]. This shows the possible applications of H – aggregates in the field of photovoltaics and non linear optics.

Fluorescence resonance energy transfer (FRET) is an electronic dynamic phenomenon that can occur through the transfer of excited state energy from donor to acceptor [15, 16]. The extent of energy transfer from donor molecule to acceptor molecule depends upon the area of spectral overlapping between the fluorescence spectrum of donor with the absorption spectrum of the acceptor, relative orientation of the donor and acceptor transition dipoles and the distance between these molecules [15, 16]. An excited molecule can transfer its energy to another identical molecule (homo transfer) or different molecule (hetero transfer) through different processes viz. radiative reabsorption or direct charge exchange (Dexter type transfer) or non radiative fluorescence resonance energy transfer (FRET) [17]. FRET is sensitive at distances 1 – 10 nm between donor and acceptor molecules whereas radiative energy transfer occurs at large distances and dexter type energy transfer occurs at close distances (<1 nm) [18]. It means to be a FRET pair two molecules have to satisfy the above mentioned conditions.

FRET is very important for the understanding of some biological systems and has potential applications in optoelectronic and thin film devices [19, 20]. Combining FRET with optical microscopy, it is possible to determine the distance between two molecules within nanometers. FRET mechanisms are also important for different other phenomenon such as photosynthesis, Brownian dynamics [21], structure and assembly detection of proteins [22], designing biosensor [23], hard water sensor [24], ion sensor [25] etc.

Different types of cyanine derivatives have been extensively studied around the world in the restricted geometry of ultrathin films prepared by LB technique for their numerous applications in different fields of research [26 – 30]. In one of our previous works we have demonstrated the control of J - aggregate formation of a thiacyanine dye in LB films by incorporating nanoclayplatelet laponite [31]. We have also shown that the J - aggregates of thiacyanine dye in LB films decays to H – aggregates and monomer upon irradiation with the monochromatic light [32].

In this paper we report the H – aggregate formation of Oxa18 LB film, which enhances in presence of Laponite. Contrary to normal nonfluorescent behavior of H – aggregates, Oxa18 H – aggregates show fluorescent behavior. Generally dye aggregation quenches the fluorescence thereby affecting the non radiative energy transfer from donor to acceptor molecule badly. However in the present case the fluorescent behavior of Oxa18 H - aggregates eventually enhance the amount of energy transfer from Oxa18 (donor) to sRhb (acceptor) molecule. Therefore, investigations on Oxa18 H – aggregate fluorescence and its effect on energy transfer is very important from the application point of view.

**2. Experimental**

3,3 – Dioctadecyloxacarbocyanine perchlorate (Oxa18) (fig. 1(a)) and Sulphorhodamine B monosodium salt (sRhb) (fig. 1(b)) purity >99%, were purchased from Sigma-Aldrich chemical company and were used as received. Working solutions for Oxa18 were prepared by dissolving Oxa18 in spectroscopic grade chloroform (SRL) or methanol (SRL) and sRhb in spectroscopic grade methanol (SRL). Here chloroform solution of Oxa18 was used for the LB film preparation and methanol solution was used for energy transfer study in the solution between Oxa18 and sRhb. The clay mineral used in the present work was Laponite, obtained from Laponite Inorganics, UK, and used as received. The size of the Laponite was less than 0.05μm, and its CEC was 0.74 meq/g, [33].

A commercially available LB film deposition instrument (Apex 2000C, India) was used for isotherm measurement and monolayer film preparation. Either pure Milli-Q water or Laponite dispersions stirred for 24 h in Milli-Q water were used as sub phase. The Laponite concentration was fixed at 2 ppm. Solutions of Oxa18 were spread on the sub phase with a micro syringe. Allowing 15 and 30 min waiting time, in case of water and Laponite dispersion respectively, the barrier was compressed at the rate of 5 mm. min$^{-1}$ to record the surface pressure–area per molecule ($\pi$ – A) isotherm. The surface pressure ($\pi$) versus average area available for one molecule (A) was measured by a Wilhelmy plate arrangement, as described elsewhere [4]. The films were found to be stable and data for $\pi$–A isotherms were acquired by a computer interfaced with the LB instrument. Each isotherm was obtained by averaging at least five runs. Monolayer films were deposited in upstroke (lifting speed 5 mm min$^{-1}$) at a desired fixed surface pressure onto fluorescence grade quartz plates for spectroscopic study of LB films. The transfer ratio was found to be 0.98 ± 0.02.

The concentration of the individual dye in the mixture solution (both in presence and absence of clay) was 4×10$^{-7}$M.

In order to study energy transfer between Oxa18 and sRhb in the ultrathin film, it is necessary to fabricate a mixed film consisting of both these two dyes. The mixed film has been fabricated by using LB – LbL bitechnique in the following way:

At first 10 bilayerd LB film of Oxa18 was fabricated. Then this film was dipped into the sRhb solution for 15 minutes to adsorb the sRhb dyes onto the Oxa18 LB film. The film was taken out and dried followed by dipping into clean water for 2 minutes. This was done in order to wash out the surplus sRhb molecules from the film surface.

Fluorescence spectra and UV-Vis absorption spectra were measured by a Perkin Elmer LS 55 spectrophotometer and Perkin Elmer Lambda 25 spectrophotometer respectively. All the measurements were performed at room temperature.

**3. Results and discussion**

*3.1. Monolayer characteristics at air – water interface*

In order to check the monolayer characteristic of Oxa18 at air water interface, we have spread 100 µl of chloroform solution (0.5 mg/ml) of Oxa18 at the air water interface. After waiting 15 minutes to evaporate the volatile solvent, the barrier was compressed very slowly and the corresponding surface pressure – area ($\pi - A$) isotherms were recorded.

Fig.2 shows the ($\pi - A$) isotherms of the monolayer films of Oxa18 on pure water subphase (fig.2, curve a) and on Laponite dispersion subphase (fig.2 curve b). Pure Oxa18 isotherm shows steep rising up to collapse pressure with an initial lift off area 1.54 nm$^2$. Here lift off area is defined as the average area per molecule when the surface pressure just starts rising i.e just above zero surface pressure. Lift off area is determined by the method described by Ras et al [34]. The stable monolayer thus formed was then transferred on to solid substrate to form mono and multilayered LB films.

On the other hand the isotherm of Oxa18 on the Laponite dispersion subphase is different from that on the pure water subphase. Here the initial lift off area shifts towards larger area per molecule to 1.74 nm$^2$. It is worthwhile to mention in this context that the cation exchange capacity (CEC) of Laponite used in this study is 0.74 meq/g. With a CEC of 0.74 meq/g and an estimated surface area of 750–850 m$^2$/g, the average area per negative charge is 1.68–1.91 nm$^2$. The observed lift-off area 1.74 nm$^2$ for Oxa18 isotherm in presence of Laponite lies in this range. Therefore, it can be concluded that every cationic dye molecule in the monolayer neutralizes one negative charge on the Laponite particles by an ion exchange reaction. The Laponite particles are fixed at the air–water interface to form a so-called hybrid monolayer containing Laponite layers and dye molecules. This hybrid monolayer was then transferred on to solid substrate to form hybrid LB films.

*3.2. Spectroscopic characterization*

Fig. 3(a) and (b) shows the UV-Vis absorption and steady state fluorescence spectra of LB film of Oxa18 lifted at different surface pressure (5,10,15,20 mN/m) along with the Oxa18 LB film in presence of Laponite lifted at 15 mN/m surface pressure and Oxa18 in chloroform solution. The Oxa18 solution absorption spectrum shows distinct bands within the 425-525 nm spectral region with a prominent 0-0 band at 494 nm along with a weak high energy band at 466 nm. Now to check whether the origin of this band is due to the presence of dimeric species or merely a vibronic component of monomer band we have plotted the ratio of the intensity of the high energy band with respect to the monomer band against concentration of the solution and we found that it increases with the increase in concentration of the solution (fig. available in supporting information). This confirms that the high energy band is due to the presence of dimeric species in the solution.

It is important to mention in this respect that generally the amphiphilic molecules forms a monolayer when they are spread onto the air – water interface. Then after compression the molecules comes closer thereby increasing the corresponding surface pressure up to the point of collapse. At the point of collapse the molecules may form microcrystalline aggregates or may form bilayer or trilayer. However in most of the cases the probability of formation of bilayer is high at very high surface pressure region. Steep rising portion of the isotherm indicate the formation of condensed/compact two dimensional monolayer (Langmuir film) at air – water interface [4]. Depending on the requirement normally Langmuir film/floating monolayer is transferred onto solid substrate at that stage (condensed stage) to form LB films. In our case the collapse pressure for both the isotherm is around 35 – 38 mN/m. So considering the above facts to avoid the chance of formation of bilayer and to have compact film we have selected the surface pressure 15mN/m for hybrid LB film deposition.

The absorption spectrum of Oxa18 in LB film is quite interesting. The absorption spectra of Oxa18 LB film have two distinct prominent bands at 460 nm and 498 nm. The monomer band of the LB film is slightly red shifted by 4 nm with respect to the solution absorption spectrum. This red shift may be due to the change in micro environment when the molecules go from solution to solid state. On the other hand high energy band at 460 nm is blue shifted by 6 nm with respect to the solution. This blue shift along with the increase in the intensity of the 460 nm band in LB film may be due to the formation of H - aggregates in the LB film.

It is relevant to mention in this context that according to Kasha – exciton theory, H – aggregates are formed when interaction between transition dipoles of two or more chromophores are arranged parallel to each other [35 - 36]. In case of H – dimer due to the interaction between two transition dipoles, the exciton band splits into two components. Only the transition to the higher excited state is allowed for parallel alignment of the transition dipoles in the H – dimer. The transition to the lower excited state is not allowed as the resulting transition moment is zero. This leads to a blue shift of the absorption maximum with respect to the monomer

It may be mentioned in this context that Ras et al (37) observed such high energy shoulder in the absorption spectrum of Oxa18 in ultrathin films and reported to be due to Oxa18 H – aggregate. There are several other works using cyanine dyes where high energy shoulder of H - dimer along with the monomeric band was also observed [8 - 9].

In order to check the effect of incorporation of Laponite on the Oxa18 H – aggregate in LB films we have taken the absorption spectrum of Oxa18 – Laponinte hybrid LB film deposited at 15 mN/m surface pressure. Here the most interesting thing is that the intensity of the H - band predominates over the monomer band and also it is more blue shifted by 3 nm than that in absence of Laponite. This result clearly indicates that the extent of aggregation increases in presence of Laponite resulting in the predominance of H - aggregate band. This increase in H - aggregated band in presence of Laponite may be due to the rearrangement of the Oxa18 molecules adsorbed onto the Laponite layers. Also from isotherm study it is already confirmed that almost all the negative charge on the Laponite surface are neutralized by adsorption of Oxa18 molecules onto the Laponite surface by ion exchange reaction, which confirms the formation of hybrid film at the air – Laponite dispersion subphase. Laponite is a negatively charged clay particle with high cation exchange capacity and surface activity. On the other hand Oxa18 is cationic. Now due to electrostatic interaction, Oxa18 molecules are adsorbed onto the Laponite particles. In the conventional pure Oxa18 LB film the driving force responsible for binding the Oxa18 molecules onto the substrate was weak Van der Walls force. But in the case of organic – inorganic hybrid film the driving force is electrostatic interaction between two oppositely charged molecules. It means the interaction scheme for holding the Oxa18 molecules onto the solid substrate is different in the hybrid film than the conventional LB film. This difference in the interaction scheme in presence of Laponite results in the rearrangement of molecular organization in the hybrid film, which may favor the formation of H – aggregate. Again due to the more stable electrostatic interaction between the Oxa18 and Laponite particles the $\pi - \pi$ stacking of the molecular aggregate increases, which may in turn increase the extent of H – aggregation in the hybrid film. It is important to recall in this context that according to exciton theory in ideal H – aggregate, molecules are stacked in plane to plane manner [35]. The absorption band is located at higher energy than monomer band and also it is non fluorescent in nature. But in some cases specially when the molecules are stacked imperfectly or the molecules possesses inclined geometry in the aggregates, in that case H – aggregate does fluoresce [6 - 12]. So in our case in presence of Laponite it may happen that the molecular organization on the Laponite surface favors the condition for imperfect stacking thereby increasing H – aggregation in the organic - inorganic hybrid film. Our later studies on emission and excitation spectra also confirm the fluorescence behavior of Oxa18 H – aggregate in hybrid LB films.

Oxa 18 solution fluorescence spectra (fig. 3 (b)) possesses distinct monomer band at 512 nm. However the LB film emission spectra of Oxa18 is very interesting. It is observed that at higher surface pressure the LB film emission spectra consist of two bands. One is the monomer band which has identical peak position to that of the solution emission maxima and the other is a weak hump situated at longer wavelength region at around 550 – 560 nm. Here the monomer band is broadened with respect to solution spectrum. This broadening may be due to change of microenvironment and aggregation when the Oxa18 molecules are transferred from solution to the restricted geometry of ultrathin films. The intensity of this longer wavelength band increases when the Oxa18 molecules are adsorbed on to the Laponite layers. The origin of this longer wavelength band may be due to the formation of excimer or the rare type of fluorescence from the H – aggregated species due to the imperfect stacking of the molecules. Now to check the origin of this longer wavelength band we have measured the excitation spectra of the Oxa18 LB film in presence of Laponite monitored at emission maxima 510 nm and 556 nm. Inset of fig. 3 (b) shows the corresponding excitation spectra. Excitation spectrum monitored at 510 nm shows prominent peak at 496 nm (fig.3 (b) inset, curve – 1) and replicates the Oxa18 monomer absorption band. On the other hand the excitation spectrum monitored at 556 nm shows prominent band with peak at around 468 nm (fig.3(b) inset, curve – 2), which replicate the Oxa18 H – band. This suggests that the origin of the longer wavelength emission band is due to the fluorescence from the Oxa18 H – aggregate. Ras et al also mentioned about the fluorescent behavior of H – aggregate of Oxa18 in ultrathin film [37]. It is important to mention in this context that fluorescence behavior of H – aggregates of various cyanine dyes have already been reported [6 – 9]. The reason for the H – aggregate fluorescence has been ascribed due to imperfect stacking of the molecules on the aggregate, or due to the inclined molecular arrangement in the aggregates [6 - 9]. In this case the dye Oxa18 has both conjugate π-electron system and hydrophobic group. Due to the presence of conjugate π-electron system charge is delocalized, which may account for the non-uniformity in the type of interaction between each individual molecular plane in the aggregate. Besides the interaction among the hydrophobic groups also helps in the tilting of molecular plane in the aggregate. These types of interactions may be responsible for the imperfect stacking or inclined geometry of Oxa18 molecules in the H-aggregate. Also there may be suppression of the nonradiative decay channels due to the increased rigidity of the molecular chain in the π-π stacked aggregate as suggested by Cigáň et al [10].

However, supra-molecular structure of fluorescent H-aggregates i.e. geometrical structure and orientation of molecules, is not fully discovered yet and is discussed based on individual ideas separately. So instead of the availability of very few data on the optical and spectroscopic properties of fluorescent H-aggregates, little is known about the structure characteristics of fluorescent H-aggregates.

Schematic diagram showing the arrangement of Oxa18 molecules in ideal H – aggregates and fluorescent H – aggregates have been shown in figure 4 (a) and 4 (b) respectively. In the present case we assume the imperfect stacking or inclined geometry of Oxa18 molecules in the fluorescent H – aggregates as shown in fig 4 (b).

*3.3. Energy transfer from Oxa18*

*3.3.1. Energy transfer in solution*

In order to study the effect of H – aggregate fluorescence on the energy transfer from Oxa18 (considering as donor) to other molecules, we have chosen sulphorhodamine B (sRhb) as acceptor and studied the energy transfer between Oxa18 to sRhb. The fluorescence spectrum of Oxa18 and absorbance spectrum of sRhb possesses sufficient overlap (inset a of fig. 5), which is a prerequisite for energy transfer to occur [15 - 16].

In order to investigate the possible energy transfer between Oxa18 and sRhb, we have measured the fluorescence spectra of Oxa18, sRhb and their mixture in both presence and absence of Laponite with excitation wavelength 480 nm. The excitation wavelength was so chosen to excite Oxa18 directly and to avoid or minimize the direct excitation of sRhb.

However one must be very much careful while choosing oppositely charged dye molecules as FRET pair due to the possibility of stable complex formation in their mixture. This will hamper the FRET process badly. Here Oxa18 and sRhb both have a π-electron system and also both of these two dyes have hydrophobic groups. The molecular structures of these two dyes are shown in fig. 1. From the structure it is clear that Oxa18 is cationic in nature and sRhb is anionic. So there must be electrostatic interaction between these two dyes in the mixture. But the nature of this electrostatic interaction is different from that of the conventional electrostatic interaction between two cationic and anionic charges where charge is localized. However in this case, as both of these two dyes possess conjugate π-electron system, delocalization of charges (resonance) occurs in these dyes. Besides this electrostatic interaction, there also exists a competing hydrophobic interaction between these two dyes in the mixture. As a combined effect of these two types of interactions, the actual interaction scheme in the present case may be a complex one rather than solely electrostatic or hydrophobic. This type of complex interaction may drive the molecules to come close enough but prevent from forming stable complex. Steric interaction between these two dyes may also prevent them to form stable complex and this is verified by our calculated value of donor (Oxa18) – acceptor (sRhb) distance, 6.89 nm as listed in table 2.

Fig. 5 shows the fluorescence spectra of pure Oxa18 (1), sRhb (2) and their mixture in methanol solution (3) as well as in clay dispersion (4). Corresponding intensity of different fluorescence bands is tabulated in table 1. It has been observed that the intensity of the fluorescence band of Oxa18 is strong and prominent where as the fluorescence band of sRhb is less intense in case of pure dye solution. The less intensity of the sRhb fluorescence band indicates very small contribution of direct excitation of sRhb molecules with excitation wavelength 480 nm. The fluorescence spectra of Oxa18 - sRhb mixture is very interesting, here the Oxa18 fluorescence intensity decreases in favor of sRhb fluorescence band. This decrease in fluorescence intensity of Oxa18 molecules is due to the energy transfer from Oxa18 to sRhb molecules. This transferred energy excites more no of sRhb molecules followed by light emission from sRhb, which is added with the original fluorescence of sRhb. As a result the intensity of sRhb fluorescence band increases. In order to confirm this we have measured the excitation spectrum for Oxa18 – sRhb mixture solution monitored at emission maxima 513 nm (Oxa18 emission maxima, fig.5 inset b curve – i) and 575 nm (sRhb emission maxima, fig.5 inset b curve – ii). Interestingly both the excitation spectra replicate the Oxa18 monomer absorbance. This confirms that the origin of sRhb fluorescence in case of Oxa18 – sRhb mixture is due to the light absorption by Oxa18 monomer and corresponding transfer to sRhb monomer.

We have also measured the fluorescence spectra of Oxa18 – sRhb mixture in clay dispersion and checked the energy transfer. Interestingly it has been observed that in presence of Laponite (fig.5, curve – 4) the fluorescence intensity of Oxa18 decreases further and sRhb intensity increases, indicating an increase in energy transfer in presence of Laponite. From table 1 it has been observed that the extent of increase in sRhb fluorescence intensity and decrease in Oxa18 fluorescence intensity in absence of Laponite is almost equal. However, in presence of Laponite dispersion the increase in sRhb fluorescence is more compared to the decrease in Oxa18 fluorescence intensity. In ideal case of energy transfer the decrease in donor fluorescence intensity should be equal to the increase in acceptor fluorescence intensity. In order to explore the reason for this mismatch we have measured the excitation spectra for Oxa18 – sRhb mixture in presence of Laponite monitored at emission maxima 513 nm (Oxa18 emission maxima, fig.5 inset b curve – iii) and 575 nm (sRhb emission maxima, fig.5 inset b curve – iv). Interestingly the excitation spectra monitored at 575 nm possess two bands with peaks at 466 nm (H – band of Oxa18 absorbance) and 496 nm (Oxa18 monomer absorption band). This indicates that fluorescence intensity of sRhb in the Oxa18 – sRhb mixture in presence of Laponite not only increased due to the energy transfer from Oxa18 monomer but also from the fluorescent Oxa18 H – aggregate band.

In order to quantify the energy transfer, we have also calculated different FRET parameters using standard Förster theory [15 - 16]. These are listed in table 2. The detail of the calculation procedure is mentioned elsewhere [38]. It has been observed that the energy transfer efficiency increases from 36.54% to 57.65 % in presence of Laponite. This is because the distance between the dyes decreases when they are adsorbed onto Laponite surfaces. Also the spectral overlap changes in presence of clay particle Laponite. This indicates that some kind of orientational change of the molecules occurred in presence of Laponite. This change in molecular orientation in presence of Laponite may give a favourable condition for the formation of Oxa18 H – aggregates.

*3.3.2. Energy transfer in ultrathin film*

In order to confirm the existence of both of these two dyes in the mixed film we have taken the absorption spectra of the mixed film. Fig.6 shows the absorption spectra of the mixed film along with the absorption spectra of Oxa18 LB film and sRhb LbL film for the reference of the corresponding peak positions. The pure sRhb LbL film shows an intense band at 570 nm and a weak high energy hump at 530 nm. The intense band at 570 nm is attributed as the monomeric band of rhodamine dye. On the other hand the band at higher energy is due to the presence of aggregated species in the film [39]. Such aggregating behavior of rhodamine dyes has already been reported by various authors [40 - 42]. On the other hand absorption characteristics of Oxa18 LB film have already been discussed in the previous section of the manuscript. It has been observed from the absorption spectra of the mixed film that it contains the peak positions attributed to Oxa18 and sRhb both. This confirms the presence of both of these two dyes in the mixed film.

Fig. 7 shows the fluorescence spectra of Oxa18 – sRhb film prepared by LB – LbL bitechnique in presence and absence of clay alongwith the pure Oxa18 LB film and sRhb LbL film.

From the figure it has been observed that energy transfer occurred for the Oxa18 – sRhb mixed films both in presence and absence of Laponite particles. The excitation spectra monitored with emission maximum 590 nm (sRhb emission maximum) in case of Oxa18 – sRhb mixed film possess both the characteristics bands of Oxa18 absorption monomer and H – aggregate (curve b, inset of fig.7). This confirms that incase of mixed films energy transfer occurs from both Oxa18 monomer and H - aggregate to sRhb monomer even in absence of Laponite. Interesting thing of our observation is that although the fluorescence intensity of Oxa18 emission band is almost same in the mixed films in absence and presence of Laponite, the fluorescence intensity of sRhb band in the mixed film in presence of Laponite is higher than that in absence of Laponite. The possible reason is may be in presence of Laponite the energy transfer from Oxa18 H - aggregate to sRhb monomer predominates.

As mentioned in previous section, the presence of Laponite particle provides a favorable condition for the Oxa18 - H aggregate formation in the ultrathin films. Also in the present case H

- aggregate does fluoresce. Therefore, fluorescent Oxa18 H - aggregate transfers energy to sRhb monomer. This results an increase in sRhb fluorescence in Oxa18 – sRhb mixed film in presence of Laponite.

## 4. Conclusion

In this paper it has been shown that Oxa18 molecules form stable Langmuir monolayer at air – water interface both in presence and absence of Laponite particles. This Langmuir film was successfully transferred onto solid support to form stable LB films. Presence of Oxa18 H – aggregate and Oxa18 monomer in ultrathin films have been confirmed by spectroscopic investigation. Incorporation of Laponite enhances the extent of H – aggregation in LB film. Contrary to normal nonfluorescent behavior of H – aggregate, Oxa18 forms fluorescent H – aggregates. This observed fluorescence may be due to the imperfect stacking or inclined geometry of Oxa18 molecules in the H – aggregates. Also there may be suppression of the nonradiative decay channels due to the increased rigidity of the molecular chain in the $\pi - \pi$ stacked aggregate. Energy transfer occurred between Oxa18 to sRhb molecules in solution and in ultrathin film. Interestingly Oxa18 H – aggregate fluorescence was found to have substantial contribution to this energy transfer.

**References**:

1. H. Kuhn, D. Moibus, H. Bucher, Techniques of Chemistry, in: A. Weissberger, B. W. Rossiter (Eds.), New York, vol. 1, Part IIIB. (1973).

2. K. Ariga, Y. Yamauchi, T. Mori, J.P. Hill, 25th Anniversary Article: What Can Be Done with the Langmuir-Blodgett Method? Recent Developments and its Critical Role in Materials Science, Adv. Mater. 25 (2013) 6477 – 6512.

3. S.A. Hussain, D. Bhattacharjee, Langmuir – Blodgett films and molecular electronics. Mod. Phys. Lett. B 23 (2009) 3437 - 3451.

4. A. Ulman, An Introduction to Ultrathin Organic Films: From Langmuir – Blodgett Films of Self Assemblies, Academic Press, New York, 1991.

5. T. Kobayashi, J-Aggregates, World Scientific, Singapore, 1996.

6. A.K. Chibisov, G.V. Zakharova, H. Gorner, Photoprocesses in dimers of thiacarbocyanines, Phys. Chem. Chem. Phys. 1 (1999) 1455–1460.

7. A.K. Chibisov, G.V. Zakharova, H. Gorner. Photoprocesses of thiamonomethinecyanine monomers and dimmers, Phys. Chem. Chem. Phys. 3 (2001) 44–49.

8. C. Peyratout, L. Daehne, Aggregation of thiacyanine derivatives on polyelectrolytes, Phys. Chem. Chem. Phys. 4 (2002) 3032–3039.

9. U. Rosch, S. Yao, R. Wortmann, F. Wurthner, Fluorescent H-aggregates of merocyanine dyes, Ange. Chem. Int. Ed. 45 (2006) 7026 – 7030.

10. M. Cigáň, J. Donovalová, V. SzÖ, J. Gašpar, K. Jakusová, and A. Gáplovsky, 7 – (Dimethylamino) coumarin – 3 – carbaldehyde and its phenylsemicarbazone: TICT excited state modulation, fluorescent H – aggregates, and preferential salvation, J. Phys. Chem. A. 117 (2013) 4870 – 4883.

11. P. Verma, and H. Pal, Unusual H – type aggregation of coumarin - 481 dye in polar organic solvents, J. Phys. Chem. A. 117 (2013) 12409 – 12418.

12. P. Verma, and H. Pal, Intriguing H – aggregates and H – dimer formation of coumarin – 481 dye in aqueous solution as evidenced from photophysical studies, J. Phys. Chem. A. 17 (2012) 4473 – 4484.

13. K. Saito, H-Aggregate formation in Squarylium Langmuir−Blodgett films, J. Phys. Chem. B. 105 (2001) 4235 – 4238

14. H.S. Zhou, T. Watanabe, A. Mito, I. Honma, K. Asai, K. Ishigure, Encapsulation of H aggregates in silica film with high nonlinear optical coefficient ($\chi^3$=3.0×10$^{-8}$ esu) by a simple sol–gel method, Mater. Lett. 57 (2002) 589 - 593.

15. T.H. Förster, Experimentelle and theoretische untersuchung des Zwis – chenmolekularen ubergangs von elektrinenanregungsenergie, Z. Naturforsh 4A (1949) 321 – 327.

16. T.H. Förster, Transfer mechanisms of electronic excitation, Diss. Faraday Soc. 27 (1959) 7 – 71.

17. S.R. Bobbara, Energy transfer between molecules in the vicinity of metal nanoparticle, PhD thesis, June, 2011.

18. D.L. Dexter, A theory of sensitized luminescence in solids, J. Chem. Phys. 21 (1953) 836.

19. G. Haran, Topical Review: Single-molecule fluorescence spectroscopy of biomolecular folding, J. Phys. Condens. Matter 15 (2003) R1291–R1317.

20. M.S. Csele, P. Engs, Fundamentals of Light and Lasers, Wiley, New York, 2004.

21. Y. Yilmaz, A. Erzan, O. Pekcan, Critical exponents and fractal dimension at the sol – gel phase transition via in – situ fluorescence experiments, Phys. Rev. E 58 (1998) 7487 – 7491.

22. B.S. Watson, T.L. Hazlett, J.F. Eccleston, C. Davis, D.M. Jameson, A.E. Johnson, Macromolecular arrangement in the aminoacyl – tRNA – elongation factor Tu – GTP ternary complex. A fluorescence energy transfer study, Biochemistry 34 (1995) 7904 – 7912.

23. D. Bhattacharjee, D. Dey, S. Chakraborty, S.A. Hussain, S. Sinha, Development of a DNA sensor using molecular logic gate, J. Biol. Phys. 39 (2013) 387-394.

24. D. Dey, D. Bhattacharjee, S. Chakraborty, S.A. Hussain, Development of hard water sensor using fluorescence resonance energy transfer, Sens. Actuators B: Chem. 184 (2013) 268– 273.

25. D. Dey, J. Saha, A.D. Roy, D. Bhattacharjee, S.A. Hussain, Development of an ion sensor using fluorescence resonance energy transfer, Sens. Actuators B: Chem. 195 (2014) 382–388.

26. M.A. Jones, P.W. Bohn, Total internal reflection fluorescence and electrocapillary investigations of adsorption at a $H_2O$−dichloroethane electrochemical interface. 1. low-frequency behavior, Anal. Chem. 72 (2000) 3776 - 3783.

27. C.F. Zhao, R. Gvishi, U. Narang, G. Ruland, P.N, Prasad, Structures, spectra, and lasing properties of new (Aminostyryl) pyridinium laser dyes, J. Phys. Chem. 100 (1996) 4526 – 4532.

28. Q. Song, Z. Xu, W. Lu, P.W. Bohn, Linear and nonlinear optical behavior in hemicyanine Langmuir—Blodgett monolayers, Colloid. Surfaces A 93 (1994) 73 - 78.

29. F. Li, L. Jin, C. Huang, J. Zheng, J. Guo, X. Zhao, T. Liu, Enhancement of second harmonic generation and photocurrent generation of a novel stilbazolium dye dimer in Langmuir−Blodgett monolayer films, Chem. Mater. 13 (2001) 192 - 196.

30. K. Rajesh, M. S. Chandra, S. Hirakawa, J. Kawamata, T.P. Radhakrishnan, Polyelectrolyte templating strategy for the fabrication of multilayer hemicyanine Langmuir − Blodgett films showing enhanced and stable Second Harmonic Generation, Langmuir 23 (2007) 8560 - 8568.

31. D. Bhattacharjee, S.A. Hussain, S. Chakraborty, R.A. Schoonheydt, Effect of nano-clay platelets on the J-aggregation of thiacyanine dye organized in Langmuir – Blodgett films: A spectroscopic investigation, Spectrochim. Acta Part A 77 (2010) 232 – 237.

32. S.A. Hussain, D. Dey, S. Chakraborty, D. Bhattacharjee, J – aggregates of thiacyanine dye organized in LB films: Effect of irradiation of light, J. Lumin. 131 (2011), 1655 – 1660.

33. T. Szabo, R. Mitea, H. Leeman, G.S. Premachandra, C.T. Johnston, M. Szekeres, I. Dekakany, R. A. Schoonheydt, Adsorption of protamine and papain proteins on saponite, Clays Clay Miner. 56 (2008) 494 – 504.

34. R.H.A. Ras, J. Nemath, C.T. Johnston, E. Demasi, I. Dekakany, R.A. Schoonheydt, Hybrid Langmuir – Blodgett monolayers containing clay minerals: effect of clay concentration and surface charge density on the film formation, Phys. Chem. Chem. Phys. 6 (2004) 4174 – 4184.

35. M. Kasha, H.R. Rawls, M.A. El – Bayoumi, The exciton model in molecular spectroscopy, Pure Appl. Chem. 11 (1965) 371 – 392.

36. A.S. Davydov, Theory of molecular excitons, McGraw-Hill, New York, 1962.

37. R.H.A. Ras, Molecular and particulate organization in organo – clay monolayers. Dissertationes De Agricultura, Katholieke Universiteit Leuven, December, 2003.

38. D. Dey, D. Bhattacharjee, S. Chakraborty, S.A. Hussain, Effect of nanoclay laponite and pH on the energy transfer between fluorescent dyes, J. Photochem. Photobiol. A: Chem. 252 (2013) 174 – 182.

39. S.A. Hussain, S. Banik, S. Chakraborty, D. Bhattacharjee, Adsorption kinetics of a fluorescent dye in a long chain fatty acid matrix, Spectrochim. Acta Part A. 79 (2011) 1642 – 1647.

40. R.W. Chambers, T. Kajiwara, D.R. Kearns, Effect of dimer formation on the electronic absorption and emission spectra of ionic dyes. Rhodamines and other common dyes, J. Phys. Chem. 78 (1974) 380 - 387.

41. K.K. Rohatgi, G.S. Singhal, Nature of bonding in dye aggregates, J. Phys. Chem. 70 (1966) 1695 – 1701.

42. J.E. Selwyn, J. Steinfeld, Aggregation of equilibriums of xanthene dyes, J. Phys. Chem. 76 (1972) 762 - 774.

**Figure captions**:

**Fig. 1.** Molecular structure of (a) Oxa18 and (b) sRhb.

**Fig. 2.** Surface pressure Vs area per molecule isotherm of Oxa18 molecule on (a) pure water sub phase (b) Laponite dispersion sub phase.

**Fig. 3.** (a) UV-Vis absorption (b) steady state fluorescence spectra of Oxa18 solution, LB film lifted at different surface pressure (5, 10, 15, 20 mN/m) and Oxa18 LB film in presence of Laponite lifted at surface pressure of 15 mN/m. The inset of fig. 2(b) shows the excitation spectra of the LB film of Oxa18 in presence of Laponite monitored at emission maxima 510 nm (curve – 1) and 556 nm (curve – 2). All the films were 10 bilayer LB films.

**Fig. 4.** Schematic diagram of (a) ideal stacking of Oxa18 H – aggregate and (b) imperfect or inclined stacking of Oxa18 H – aggregate.

**Fig. 5.** Fluorescence spectra of pure Oxa18 (1), sRhb (2) and Oxa18 – sRhb mixture in absence (curve – 3) and presence of Laponite (curve – 4). The fluorescence spectra were measured with excitation wavelength 480 nm. Inset (a) shows the spectral overlap between the Oxa18 fluorescence spectrum and sRhb absorbance spectrum. Inset (b) shows the excitation spectra monitored at donor (Oxa18) emission maxima 513 nm for Oxa18 – sRhb mixture solution in absence of clay (curve – i), in presence of clay (curve – iii) and at acceptor (sRhb) emission maxima 575 nm for Oxa18 – sRhb mixture solution in absence of clay (curve – ii), in presence of clay (curve – iv).

**Fig. 6.** Absorption spectra of pure Oxa 18 LB film, sRhb LbL film and mixture of Oxa18 - sRhb mixed film prepared by LB - LbL bitechnique. Each individual graph is normalized by dividing their absorbances by its absorption maximum so that the Y – axis lies within 0 to 1. This helped to plot all the curves in the same scale. This has been done as we have nothing to do with their individual absorbance, but to deal with their individual peak positions. Again the scale of absorbance of each individual graph was different. Therefore for comparison and better understanding all the graphs were normalized and presented in the same scale.

**Fig. 7.** Fluorescence spectra of Oxa18 LB film (curve – 1), sRhb LbL film (curve – 2), Oxa18 - sRhb mixed film without Laponite (curve – 3) and Oxa18 - sRhb mixed film with Laponite (curve – 4). All the fluorescence spectra were measured with excitation wavelength 480 nm. Inset shows the excitation spectra of the films monitored at donor emission maxima 512 nm (curve – a, in case of pure Oxa18 LB film), acceptor emission maxima 590 nm in absence of Laponite (curve – b, in case of Oxa18 – sRhb mixed film) and in presence of Laponite (curve – c, in case of Oxa18 – sRhb mixed hybrid film).

**Table 1.** Quantitative parameters describing fluorescence of pure donor ($F_D$), donor in presence of acceptor ($F_{DA}$), pure acceptor ($F_A$) acceptor in presence of donor ($F_{AD}$) in solution as collected from fig.3.

**Table 2.** Values of spectral overlap integral ($J(\lambda)$), energy transfer efficiency (E (%)), Förster radius ($R_0$) and donor – acceptor distance (r) calculated from the spectral characteristics of fig. 3.

| | Fluorescence intensity of pure donor ($F_D$) | Fluorescence intensity of donor in presence of acceptor ($F_{DA}$) | Difference ($F_D - F_{DA}$) | Fluorescence intensity of pure acceptor ($F_A$) | Fluorescence intensity of acceptor in presence of donor ($F_{AD}$) | Difference ($F_A - F_{AD}$) |
|---|---|---|---|---|---|---|
| Without clay | 792 | 505 | 287 | 50 | 337 | 287 |
| With clay | 740 | 337 | 403 | 45 | 630 | 585 |

**Table 1**

| | $J(\lambda) \times 10^{15}$ $M^{-1}$ $cm^{-1}$ $nm^4$ | E (%) | $R_0$ (nm) | r (nm) |
|---|---|---|---|---|
| Without clay | 7.54 | 36.54 | 6.29 | 6.89 |
| With clay | 9.21 | 57.65 | 6.9 | 5.85 |

**Table 2**

**Figures:**

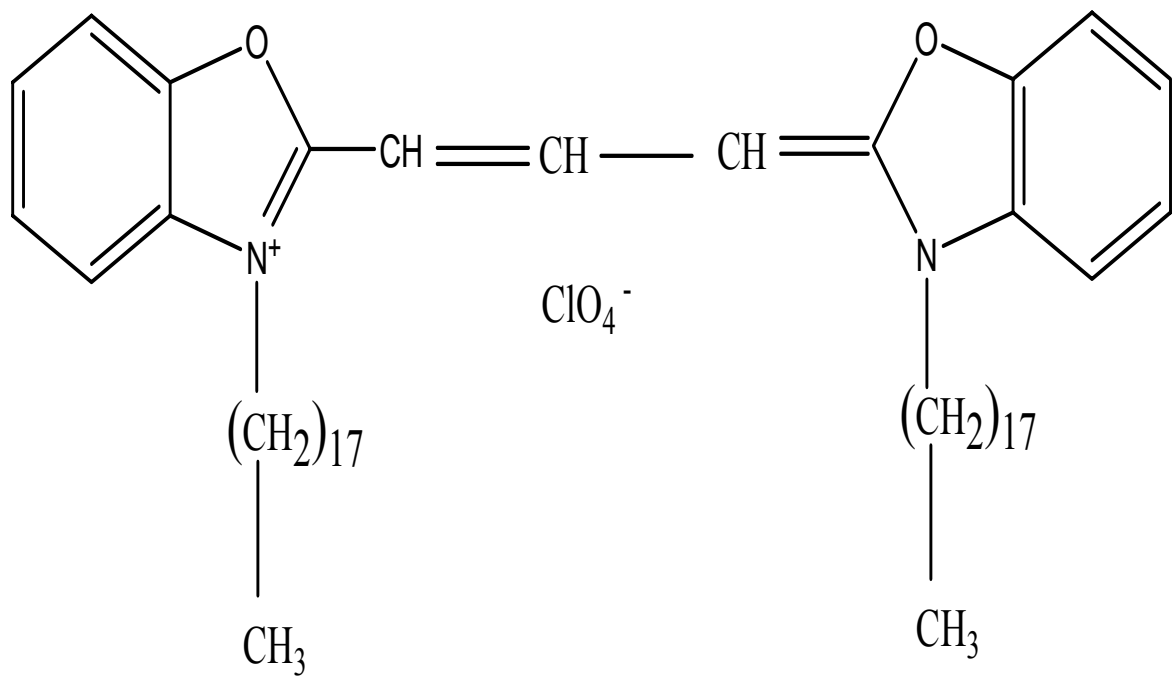

Fig. 1(a)

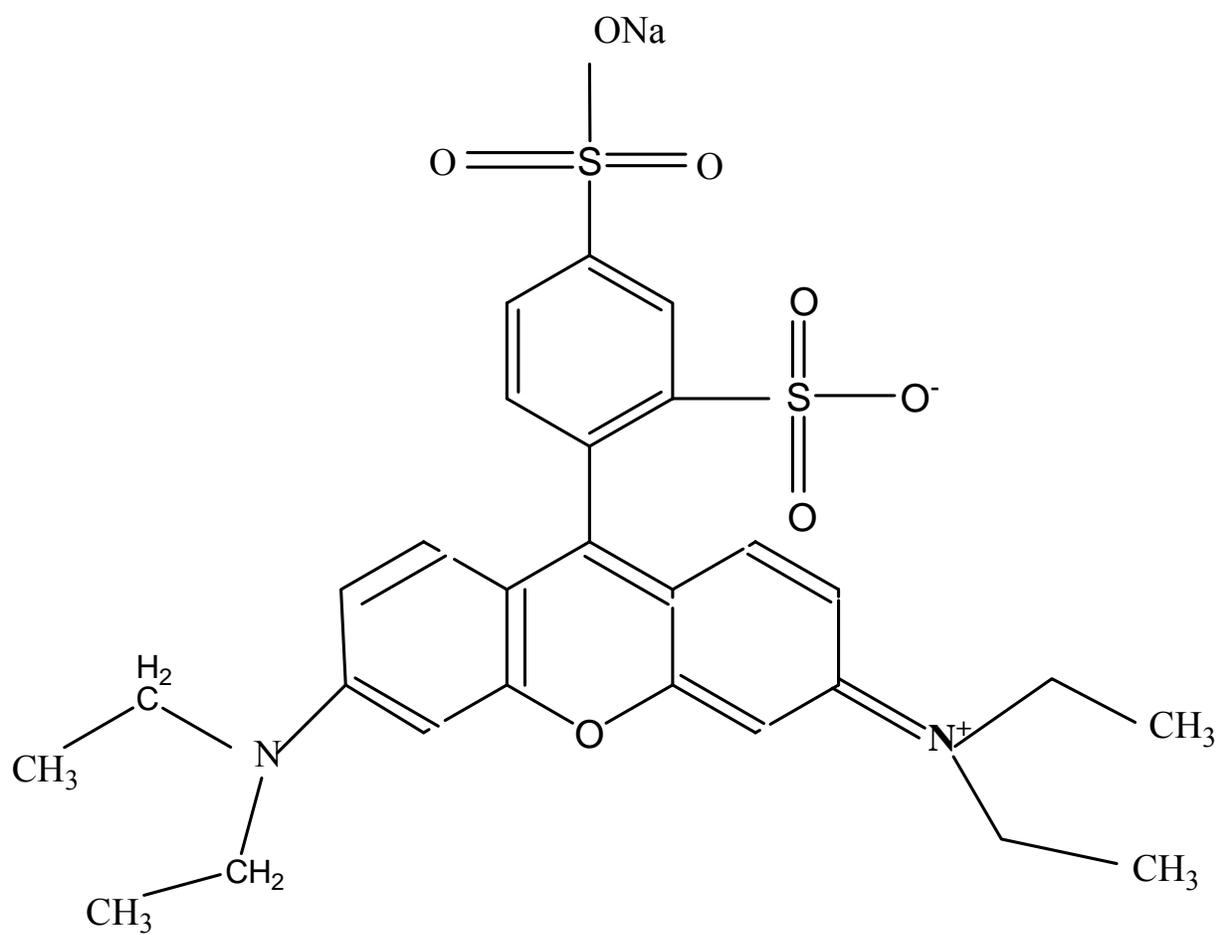

**Fig. 1(b)**

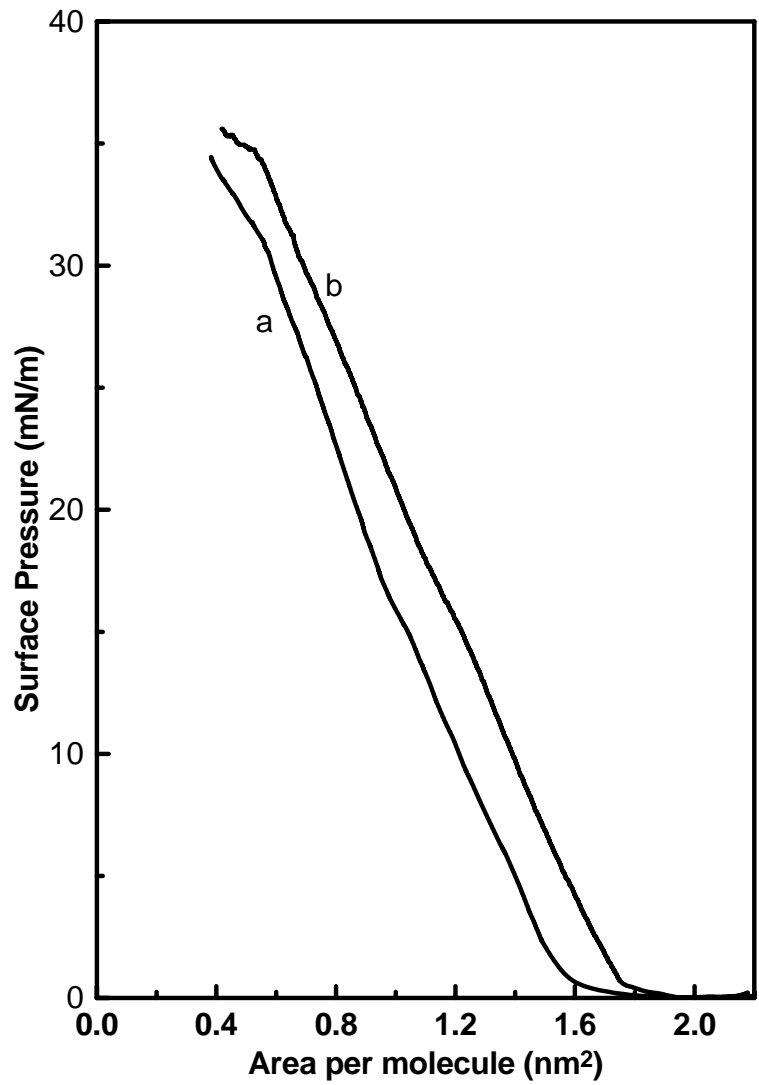

**Fig. 2**

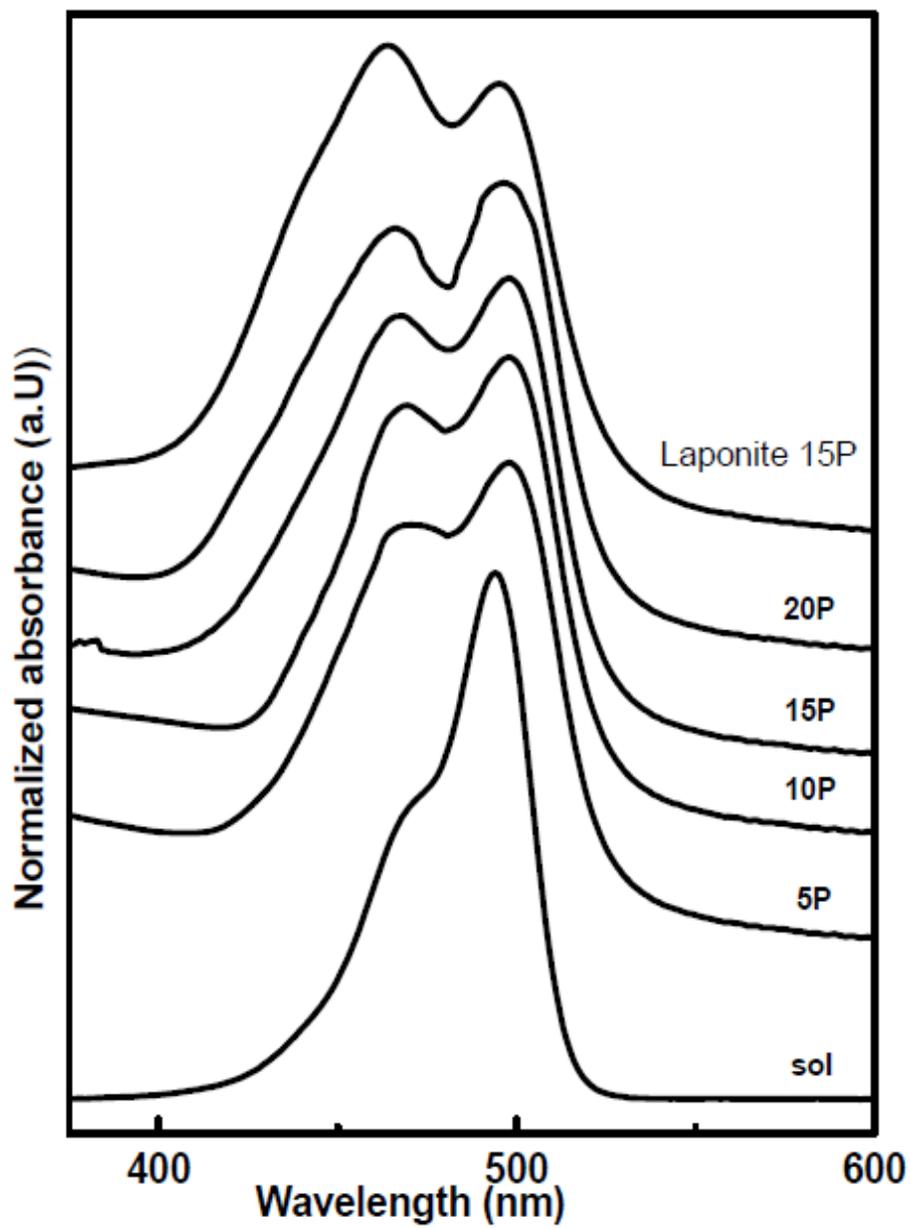

Fig. 3(a)

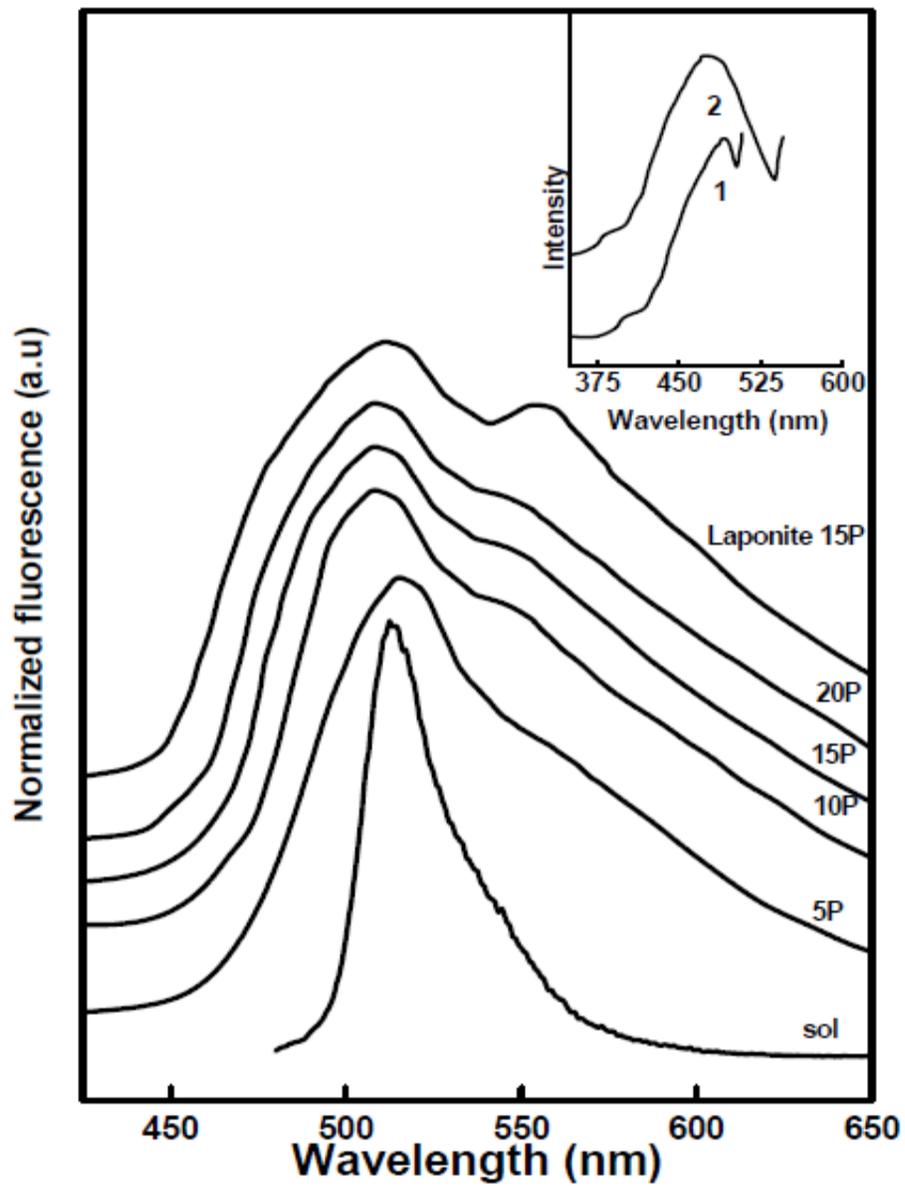

**Fig. 3(b)**

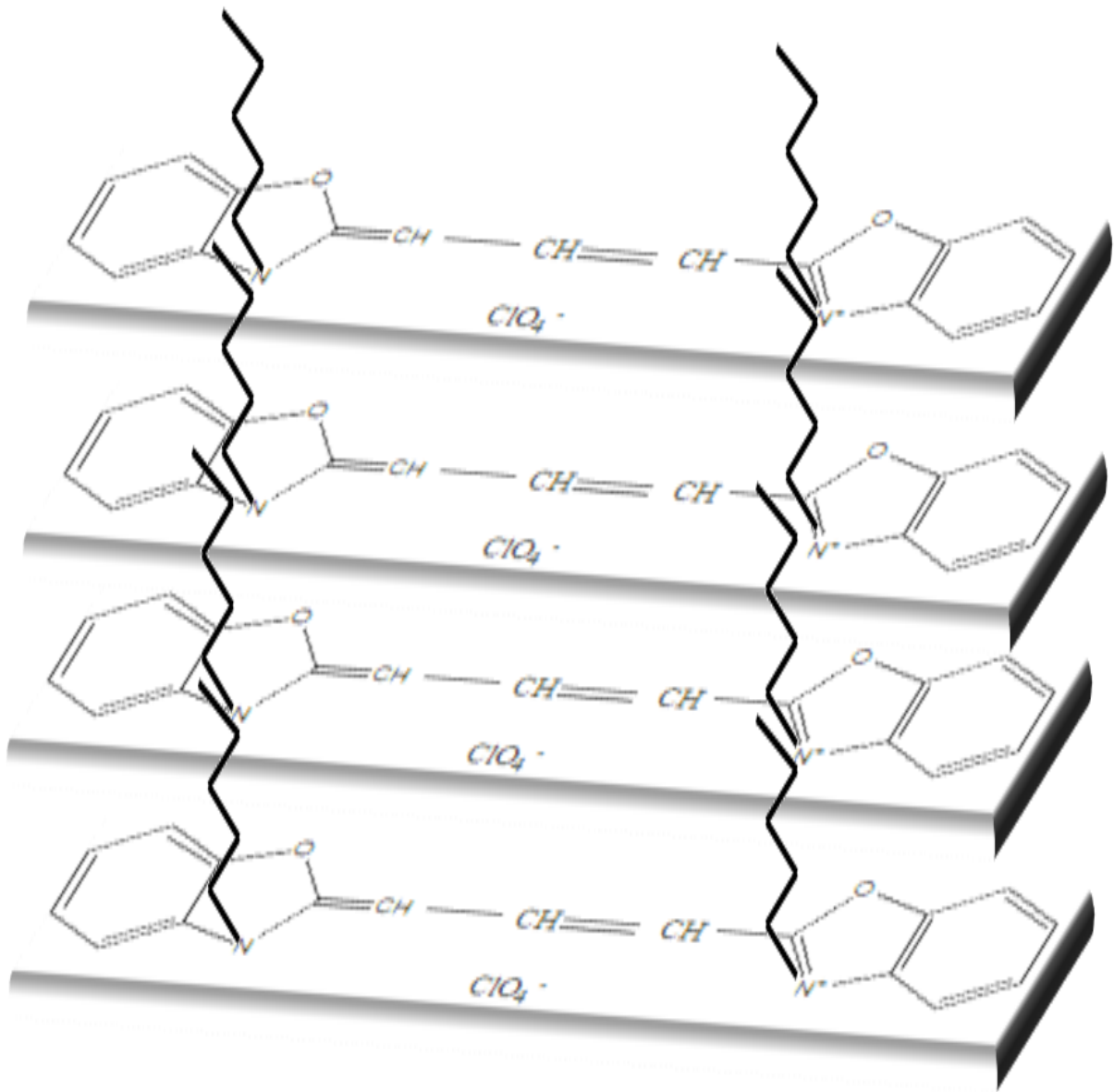

**Fig. 4 (a)**

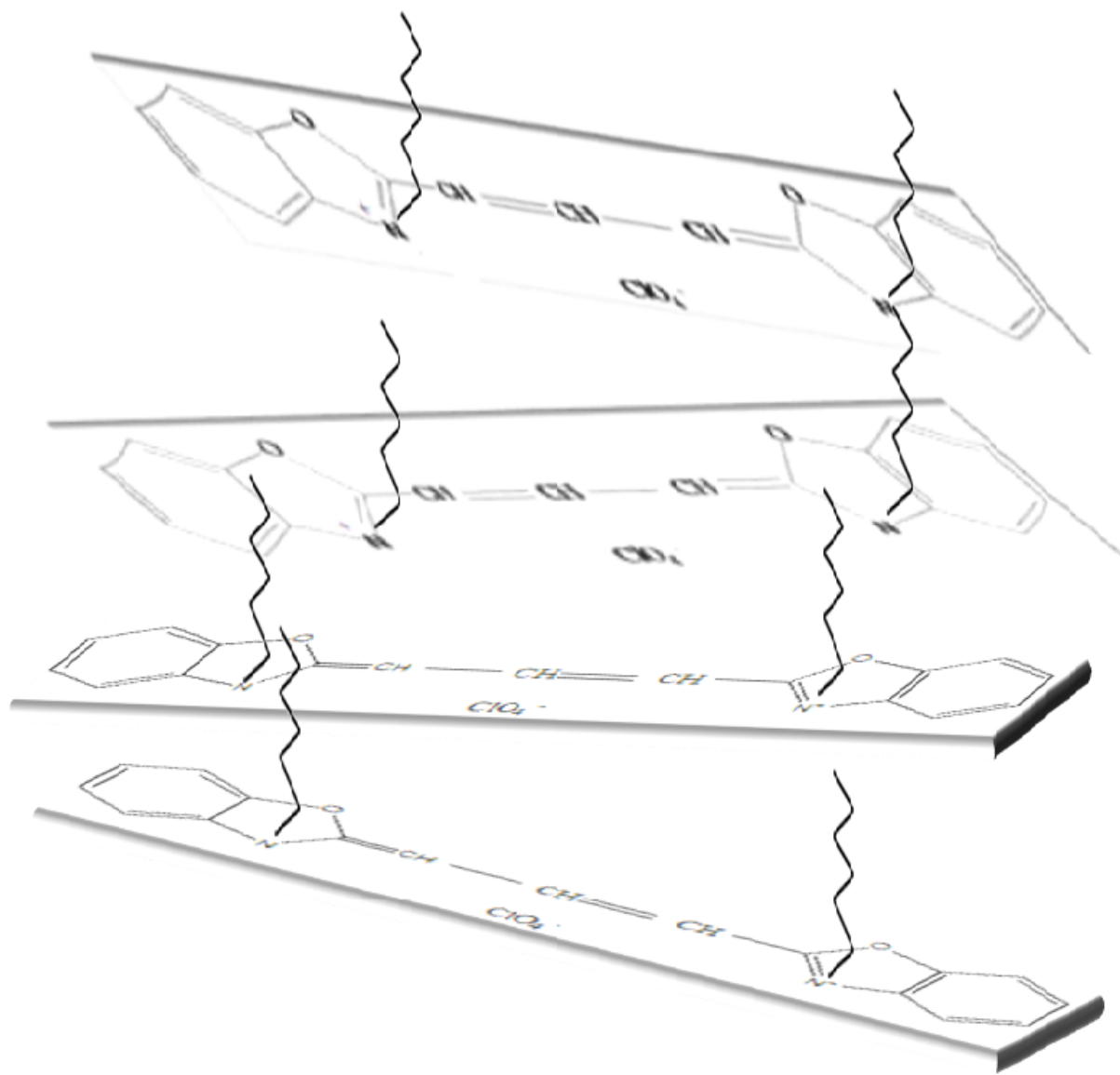

**Fig. 4 (b)**

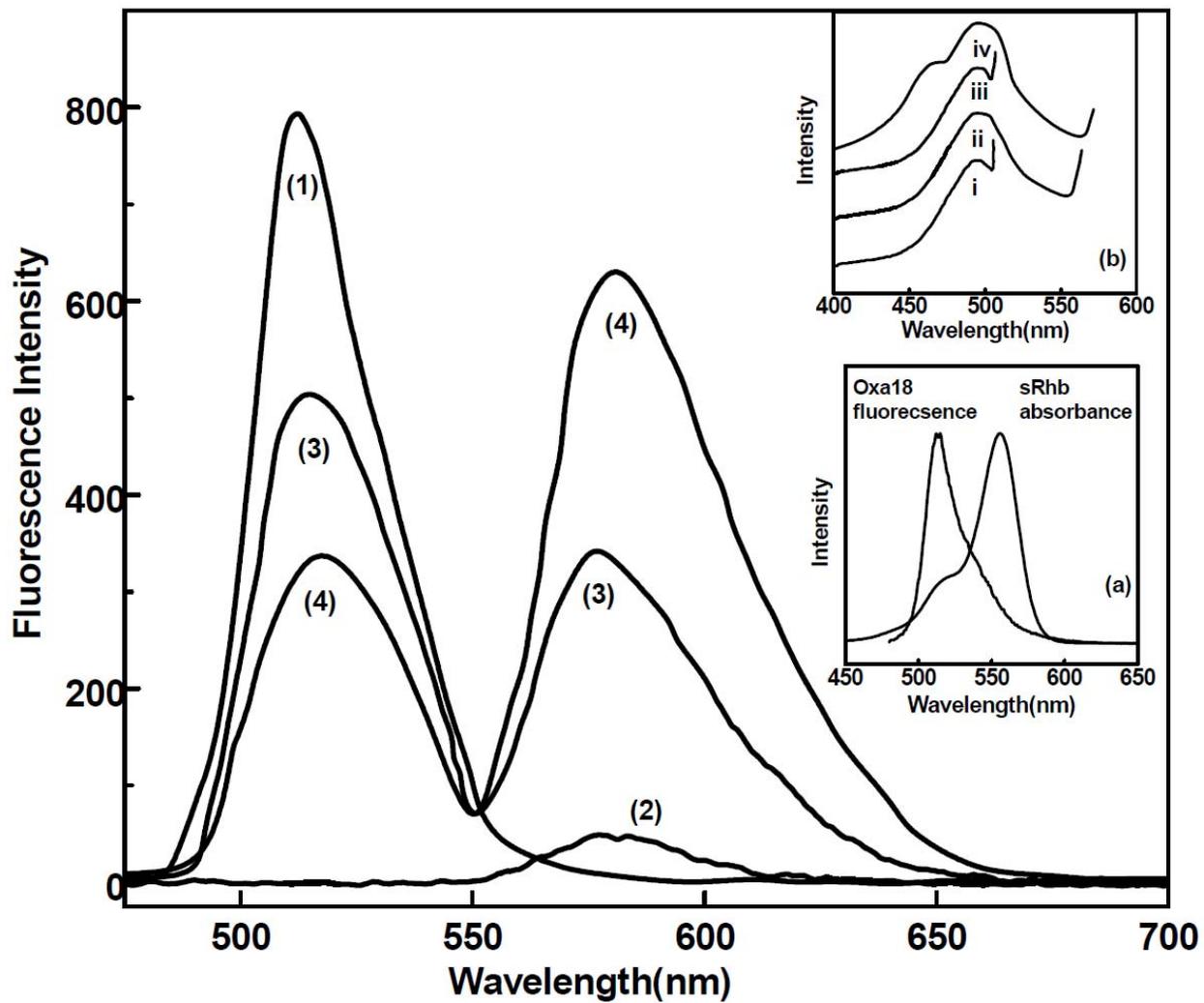

Fig. 5

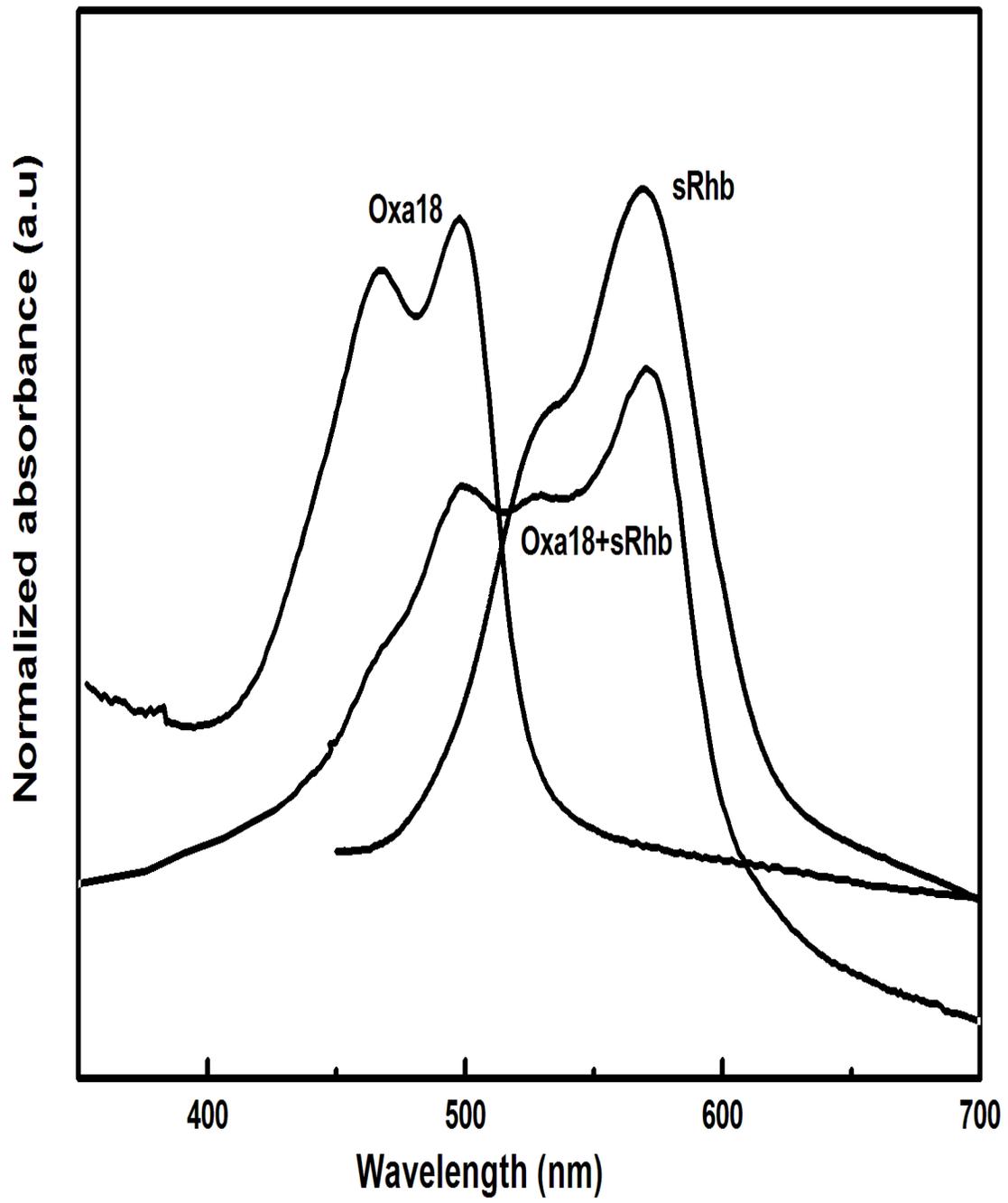

**Fig. 6**

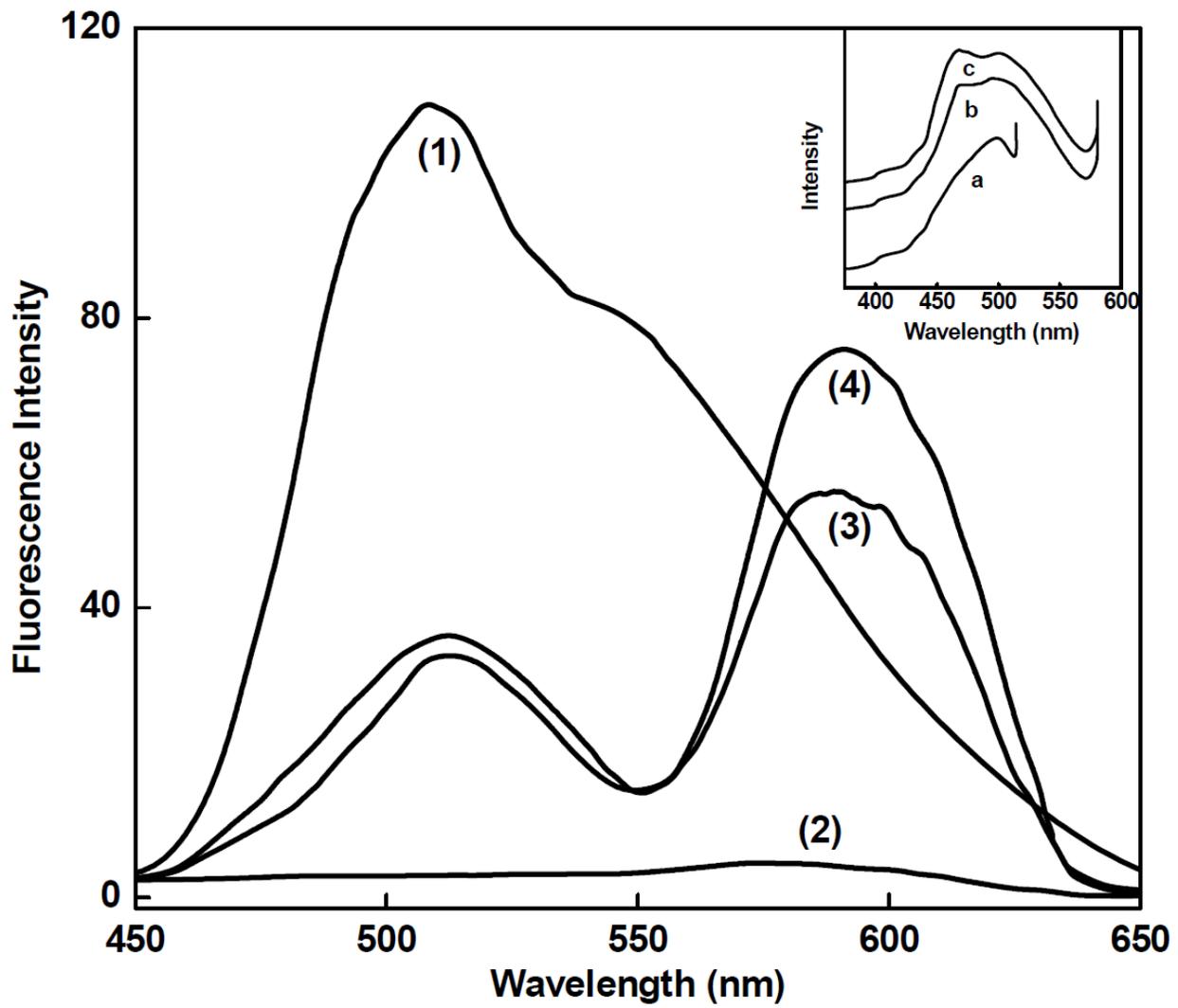

**Fig. 7**